# An Assessment of Commonly Used Equivalent Circuit Models for Corrosion Analysis: A Bayesian Approach to Electrochemical Impedance Spectroscopy

*Runze Zhang[a], Debashish Sur[b,c], Kangming Li[a], Julia Witt[d], Robert Black[e], Alexander Whittingham[e], John R. Scully[b,c], Jason Hattrick-Simpers[a*]*

   a. Department of Material Science and Engineering, University of Toronto, Toronto, Ontario, Canada

   b. Center for Electrochemical Science and Engineering, University of Virginia, Charlottesville, VA USA 22904

   c. Department of Materials Science and Engineering, University of Virginia, Charlottesville, VA USA 22904

   d. Division of Material and Surface Technologies, Federal Institute of Materials Research and Testing (BAM), 12205 Berlin, Germany

   e. Clean Energy Innovation Research Centre (CEI), National Research Council Canada, Mississauga, Ontario, Canada

Email: jason.hattrick.simpers@utoronto.ca

**Abstract**:

Electrochemical Impedance Spectroscopy (EIS) is a crucial technique for assessing corrosion of a metallic materials. The analysis of EIS hinges on the selection of an appropriate equivalent circuit model (ECM) that accurately characterizes the system under study. In this work, we systematically examined the applicability of three commonly used ECMs across several typical material degradation scenarios. By applying Bayesian Inference to simulated corrosion EIS data, we assessed the suitability of these ECMs under different corrosion conditions and identified regions where the EIS data lacks sufficient information to statistically substantiate the ECM structure. Additionally, we posit that the traditional approach to EIS analysis, which often requires measurements to very low frequencies, might not be always necessary to correctly model the appropriate ECM. Our study assesses the impact of omitting data from low to medium-frequency ranges on inference results and reveals that a significant portion of low-frequency measurements can be excluded without substantially compromising the accuracy of extracting system parameters. Further, we propose simple checks to the posterior distributions of the ECM components and posterior predictions, which can be used to quantitatively evaluate the suitability of a particular ECM and the minimum frequency required to be measured. This framework points to a pathway for expediting EIS acquisition by intelligently reducing low-frequency data collection and permitting on-the-fly EIS measurements.

**Introduction**:

In an era where the need grows for advanced materials to meet increasing social demands, material reliability has become ever more important[1,2]. Material degradation, especially in the form of aqueous corrosion, undermines the integrity and longevity of equipment and infrastructure, leading to substantial financial burdens and raising safety concerns[3–7]. To mitigate material deterioration and failure, it is important to understand the transport mechanisms during electrochemical corrosion to develop design rules for corrosion resistant materials[8]. In this context, Electrochemical Impedance Spectroscopy (EIS) is a cornerstone electrochemical technique widely used to diagnose and analyze different corrosion processes[9–11], such as pitting corrosion in stainless steels[12], aqueous passivation[13], and crevice corrosion in marine environments[14]. It offers a sensitive and non-destructive method to quantify kinetic parameters and elucidate corrosion mechanisms[15,16]. However, the interpretation of EIS relies on the selection of an appropriate equivalent circuit model (ECM) that accurately describes the electrochemical processes occurring within the system, which can be challenging and time-consuming[10,17].

One primary obstacle lies in the complex task of developing accurate models for electrochemical systems. The success of designing a suitable ECM depends on having a good grasp of the system's dynamics, which permits the selection of circuit elements that sufficiently describe the underlying processes[18]. The chosen ECM must balance complexity to capture the essential electrochemical phenomena while avoiding overfitting the data[19]. Different corrosion mechanisms, ranging from uniform corrosion to localized pitting or crevice corrosion – necessitate distinct ECMs[20,21]. For instance, a basic Randle's circuit may suffice for elementary metal oxidation processes on a planar electrode surface, whereas more complex scenarios demand intricate models with additional elements[22], like a Warburg element to account for diffusion processes[23].

The community has developed and adopted various ECMs for common corrosion scenarios, such as the simple Randle circuit described above or a series circuit with two Randle elements for corroding systems with two double layers[21,22]. However, there is potentially a lack of alignment between the complexity of the standard ECMs used to fit the EIS data and the actual information contained within the data. It can be the case that despite the perceived physical appropriateness of the ECM to the material system, the contribution from certain elements of the circuit to the EIS data may not be sufficient to reliably infer their value. In such a case, Occam's Razor would suggest that a simpler ECM be used to mitigate the possibility of overfitting, but to date there have not been many studies providing quantitative metrics for this evaluation. This highlights the need to critically assess the suitability of conventional ECMs for corrosion analysis based on the signal provided within the data[24].

Beyond the complexities of ECM selection, EIS analysis is often measured down to very low-frequency (e.g. 10^-3 Hz) to safeguard against missing important phenomena governing or playing a string role in the corrosion process. These measurements significantly prolong experimental acquisition times; for instance, a single measurement at 10^-3 Hz requires a minimum of approximately 0.3 hours, thereby imposing practical constraints on experimental throughput[25]. Furthermore, low-frequency measurements are inherently more susceptible to noise and drift. This can significantly reduce signal-to-noise ratio, impede reliable analysis, and require repeated measurements, significantly increasing analysis times, sometimes by hours[10,26]. On the other hand, given an ECM, the AC signal and impedance thereof reflects each component distributed across a wide frequency range[27,28]. The only time when this is not the case is at frequencies where the impedance parameter equates infinite or zero impedance. In this case information may not be contained in the AC response. This suggests that there might be an unexplored opportunity to optimize EIS measurement protocols in some cases by reducing the number of low-frequency measurements, while still preserving the ability to quantitatively infer all ECM elements. Recently, there has been emerging research on utilizing pre-trained neural networks to predict low-frequency characteristics based on high and medium-frequency EIS data[29,30]. While this approach shows promise, achieving accurate predictions typically requires a large amount of training data. Additionally, the method may face challenges when applied to out-of-distribution data[31]. A preferable alternative would be physics-informed and would be able to make its predictions without extensive training.

Bayesian inference (BI) is a powerful tool to mitigate the risk of model misspecification when inferring signal distributions from limited datasets, making it suitable to address the challenges in EIS analysis identified above[32–34]. Its probabilistic framework allows for a comprehensive evaluation of various candidate ECMs, enabling the quantification of their plausibility through statistical metrics such as posterior probabilities, posterior predictions, and Bayesian Information Criterion[35]. Our previous research has demonstrated BI's effectiveness in distinguishing between candidate ECMs and identifying the most statistically plausible models[33].

Here we focus on three commonly used ECMs in the corrosion community. The ECMs were selected to have increasing complexity, ranging from the simplest: charge-transfer controlled corrosion, metallic substrate with a passive film that regulates corrosion to more sophisticated systems, including an organic coat functioning as a barrier to corrosion. By varying the values of circuit components within their physically plausible ranges and applying BI to the resulting EIS data, we were able to monitor the quality of the ECM component posterior distributions as a function of the component values. This allows us to identify the regions beyond which the EIS data lacks sufficient information to substantiate the given ECM structure. Furthermore, we extended this analysis to determine the minimum measurement frequency at which BI can confidently infer the ECM component values. We propose that these boundaries define the frequency ranges within which the respective ECMs can be used with confidence, given the EIS data. Using this information, we suggest that it is possible to strategically reduce the number of low frequency EIS measurements taken on a well-defined system without compromising analysis accuracy. These insights provide a path toward improving the reliable usage of ECMs reported in the literature and can significantly reduce the time needed to collect EIS data for corrosion analysis. Beyond recommending specific ECM selections, this study introduces a framework designed to streamline EIS collection, enabling autonomous measurement systems that utilize EIS.

**Methodologies**:

*Corrosion System Selection* – The selected corrosion systems and their associated ECMs[36,37] are shown in Fig 1. Here, to account for the discrepancy between idealized electrochemical reactions and actual reactions, constant phase elements (CPEs) are used instead of pure capacitors[38,39]. The CPE parameters are defined as Q for the magnitude and α for the deviation from perfect capacitance according to the formula proposed by Brug et al.[40]. The first system represents uniform corrosion on a homogeneous surface that is primarily under charge transfer control. It can be modeled using an ohmic resistor ($R_s$) that represents the solution resistance, in series with a Randle element (composed of a charge transfer resistor $R_{ct}$ and a double-layer CPE $C_{dl}$) as illustrated in Fig. 1a. Fig 1b shows the model used to describe the passivation behavior of alloys in aqueous conditions, where both charge transfer and migration/diffusion processes occur. The associated ECM for the system is similar to that of the first but includes an additional Warburg element ($Z_w$) connected in series with the charge transfer resistor ($R_{ct}$) to accommodate the diffusion process. The third system is used for barrier type coated metals containing a penetrating defect exposed to corrosive electrolytes. Its corresponding ECM (Fig. 1c) includes a coating CPE ($C_c$), pore resistance ($R_p$), double-layer CPE ($C_{dl}$), charge-transfer resistance ($R_{ct}$), and a Warburg element ($Z_w$).

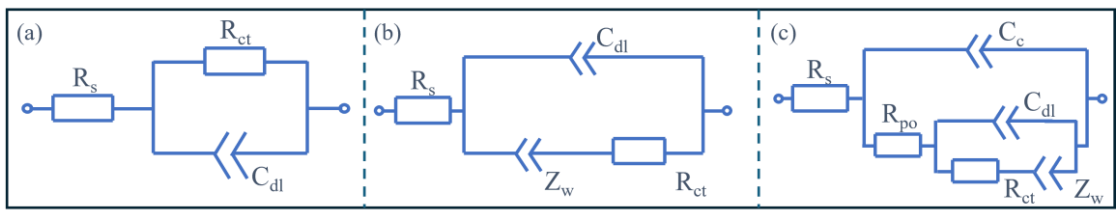

Fig 1. The commonly used Equivalent Circuit Models for three typical metallic aqueous corrosion systems. (a) Uniform corrosion (b) Passive film formation under steady state conditions (c) Corrosion of coated metallic substrate

*Generation of the Simulated EIS Data* – Here, we fixed the ECM structures described above and systematically adjusted selected component values within their physically justifiable ranges via orthogonal sampling. Three parameters in each system were selected for adjustment. For the simplest corrosion system, the key parameters varied were the charge transfer resistance ($R_{ct}$) and double-layer CPE ($C_{dl-Q}$ and $C_{dl-\alpha}$). The charge transfer resistance ($R_{ct}$) was confined to a range of $1.0 * 10^5$ to $1.0 * 10^7$ Ω, the double-layer CPE ($C_{dl-Q}$) was confined to a range of $1.0 * 10^5$ to $1.0 * 10^7$ Ω·s$^{-\alpha}$ (α refers to $C_{dl-\alpha}$), and $C_{dl-\alpha}$ was kept between 0.5 and 1.0. In our investigation of the passivating system, we adjusted the magnitude of the double-layer CPE ($C_{dl-Q}$), charge-transfer resistance ($R_{ct}$) in conjunction with the Warburg element ($Z_w$). The component ranges set for these adjustments remained unchanged as the former system, as they encompass commonly accepted ranges for each reaction parameter in corrosion analysis. When analyzing the coated metal corrosion system, we specifically adjusted the coating capacitance

($C_c$), pore resistance ($R_{po}$) and the magnitude of the double-layer CPE ($C_{dl-Q}$), within the following ranges: $1.0 * 10^7$ to $1.0 * 10^9$ $\Omega \cdot s^{-\alpha}$, $1.0 * 10^4$ to $1.0 * 10^6$ $\Omega$ and $1.0 * 10^5$ to $1.0 * 10^7$ $\Omega \cdot s^{-\alpha}$, respectively. We then infused a 2% Gaussian noise to each circuit component for every model and generated a total of 1000 simulated EIS datasets for each system across a frequency range from $10^{-3}$ Hz to $10^5$ Hz[25], and the frequencies were linearly sampled. The noise was chosen to replicate the uncertainty of reaction parameters derived from EIS measurements[26], and was selected to reflect the upper limit of noise commonly encountered in real-world scenarios.

*Model Evaluations* – Next, we examined the plausibility of the ECMs within the generated space using BI. Since the extraction of the system's ohmic resistance is straightforward, we excluded the ohmic resistor from the evaluation workflow for simplicity. We established the criterion that each ECM component must be irreplaceable and accurately inferred with high confidence from its contribution to the EIS data in order that an ECM be accepted as statistically plausible. The presence of significant divergences (over 0.1% percent of the sampling steps) or irregular/broad posterior component distributions in the inference results were taken as an indication of ECM implausibility. Each divergence suggests difficulties in thoroughly exploring the posterior distribution during the inference process[33,41,42]. We defined EIS-ECM pairs that yield narrow, well-defined near-normal probability distributions for the circuit components as being statistically plausible.

*Minimum Frequency Exploration* – To investigate the minimum frequency necessary to generate statistically plausible inference, we selected representative EIS-ECM pairs identified above and performed BI analysis as the low frequency data was dropped. Representative data were selected via human labelling based on impedance characteristics. We sequentially removed the frequency measurements for each selected EIS data two points at a time, starting from the lowest frequency. After each additional pair of points was dropped, we applied BI to track the changes in the inferred probability distributions of circuit components. To determine the minimum usable frequency, we established an accuracy threshold on the inference of ECM components of 4% relative error between the ground truth values and the mean of the posterior distributions. Once the relative error exceeds 4%, removing additional frequencies negatively impacts the integrity of the EIS analysis. Upon further reduction of the low-frequency data, a threshold frequency is reached, referred to here as the minimum inferable frequency. Beyond this frequency the posterior distributions for the ECM components exhibit divergences and lose their Gaussian shapes. This process separates the BI results into three regions, a region for which ECM components can be inferred with confidence and accuracy, a region for which they are inferred confidently but not accurately and a region for which they can neither be inferred accurately nor confidently.

**Results & Discussions:**

**Charge-Transfer Controlled Uniform Corrosion Analysis**

Firstly, we investigated the simplest homogeneous surface corrosion system, represented by a series combination of an ohmic resistor and a Randle element. By adjusting the selected ECM components associated with system's charge transfer parameters and applying BI to the EIS dataset, we identified a boundary that separates areas where BI successfully infers the model from areas where it does not, as shown in Fig 2. The results showed that a larger value of the charge transfer constant CPE ($C_{dl-\alpha}$) and a smaller charge transfer resistance ($R_{ct}$) significantly benefit the model's inferability.

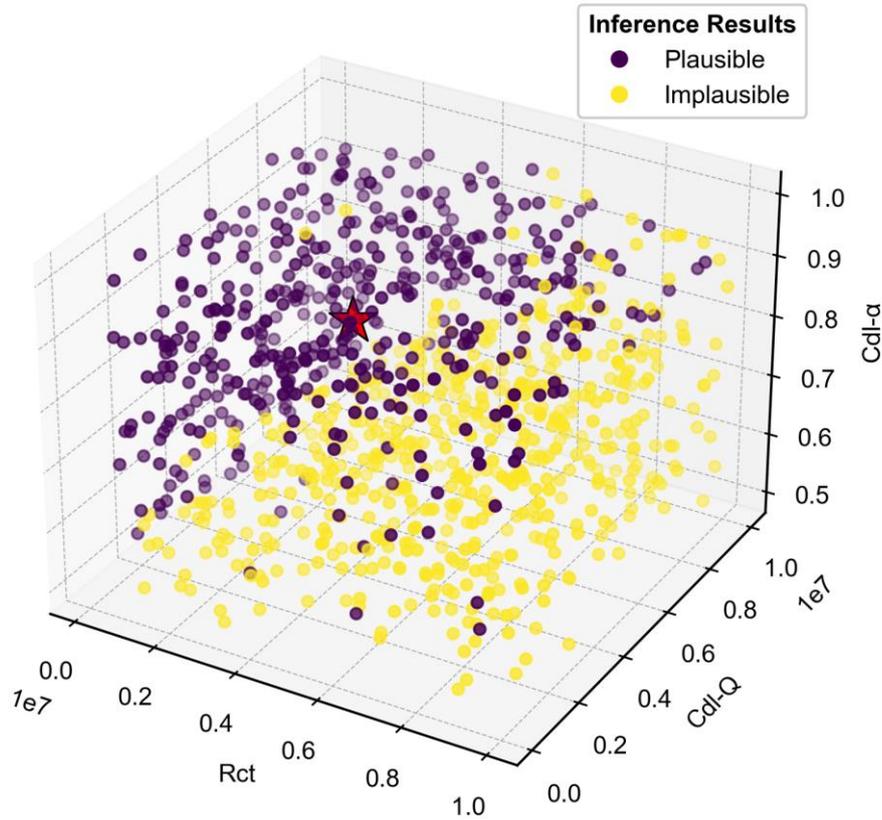

Figure 2. Distribution of statistical plausibility across EIS data with varying reaction parameters for the simplest uniform corrosion system. Purple dots: exhibiting well-distributed posterior distributions for all circuit elements; Yellow dots: containing poorly distributed posterior distribution for certain elements. $R_{ct}$: charge transfer resistance, $C_{dl-Q}$: magnitude of charge transfer CPE, $C_{dl-\alpha}$: constant of charge transfer CPE. Red mark: the selected EIS data used in the subsequent frequency reduction analysis.

We then used the parameters as input features and the BI results as objectives to build a decision tree and used the tree diagram to better identify the regions of plausibility and implausibility. The model identified the constant ($C_{dl-\alpha}$) of the charge transfer CPE as the most critical factor, with over 50% of the overall feature importance. As $C_{dl-\alpha}$ decreases, the capacitive behavior of a CPE as the dominant impedance gradually transfers to the parallel resistor with a smaller fixed resistance. The resistor has a frequency independence to its impedance and therefore the impedance exhibits less response to the frequency changes. This is observed to decrease the plausibility of the given ECM. The second most influential parameter is the value of the charge transfer resistance ($R_{ct}$), with a lower $R_{ct}$ being more conducive to a successful inference. The decision tree results, as shown in Fig. S1, show that a high charge transfer constant ($C_{dl-\alpha} > 0.82$) combined with a lower charge transfer resistance ($R_{ct} < 7.37 * 10^6$ Ω) correlates with successful signal inference. Within this range, around 96.0% of EIS data exhibited statistically plausible inference, indicating that the EIS data can be confidently fit the with associated ECM. By further constraining the value of $C_{dl-\alpha}$ to be within 0.85 to 0.98, the success rate increased to 100.0%.

Near the boundaries defined by the decision tree, there are examples of successful and unsuccessful inferences for the ECM components, where one should use with caution. In the regions for which the value of $C_{dl-\alpha}$ is smaller than 0.76 and $R_{ct}$ is larger than $2.14 * 10^6$ Ω, the BI exhibits a failure rate in excess of 93.6%. Although there are some examples of successful inference within this region, it is recommended to use the ECM carefully as the inference on the EIS is sensitive to changes in the component values. It might be the case that, despite achieving a good fit using the ECM, the reaction signals contained in the data are not strong enough to support a precise parameter estimation upon the ECM.

The impact of reducing the minimum measured EIS frequency on the inference of $C_{dl-\alpha}$ is shown in Fig. 3. The results demonstrate that we are able to decrease the number of low-frequency measurements taken without degrading the inference up to a threshold frequency. The BI inference is accurate and confident even for a minimum frequency of 0.36 Hz. Fig. S2 demonstrates that this holds for all the considered components. Beyond this range the relative error from the mean ground truth value exceeds 4%. This could be considered within experimental error depending on the user's judgment, it is therefore even feasible to exclude additional frequency data at the expense of greater component value uncertainty. As we continue to drop the low-frequency measurements, the mean values of the inferred components begin to deviate more significantly from the ground truth and divergences emerge. Eventually, for frequencies greater than 31.6 Hz, the circuit structure becomes non-inferable, containing circuit elements with implausible posterior distributions.

These observations suggest that when performing subsequent EIS measurements during an in-situ degradation study, users would only need to measure the full EIS for the initial state. For subsequent measurements they could skip the low-frequency measurements (< 0.36 Hz) and achieve a significant reduction in measurement time (over 99.9%) without detrimentally impacting the integrity of the data analysis. Clearly, the minimum frequency will also be affected by the noise level associated with the EIS measurements, however the proposed method automatically takes the noise into account during its inference. As illustrated in the supplemental materials, a lower noise level allows users to omit even more low-frequency measurements through this framework while ensuring the same analysis integrity.

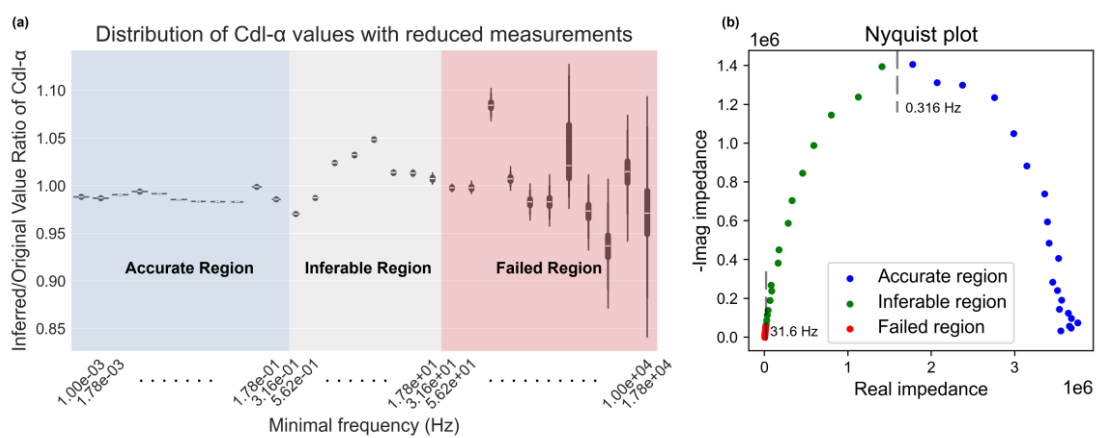

Figure 3. Impact of reducing low-frequency measurements on the inference accuracy for EIS data of the uniform corrosion system. (a) Violin plot showing the variation in the selected circuit component value ($C_{dl-\alpha}$) as low-frequency measurements are progressively excluded. (b) Nyquist plot illustrating the minimal frequency for regions with different statistical plausibility.

**Passivation of Alloys Under Aqueous Condition**

Subsequently, we focused on alloy passivation under aqueous conditions, typically modeled using a Randle circuit with an additional Warburg element to account for a diffusion process. Within the adjusted parameter ranges, we found that most of the generated EIS data failed to yield plausible inference, with only a few successful cases well characterized by a relatively small Warburg impedance ($Z_w$), as depicted in Fig. 4. Failed inferences were identified by straight lines in their Nyquist plots, indicating insufficient signal strength for accurate component value inference. This occurs because as $Z_w$ increases, the signal in the absolute sense becomes smaller, but signal from the diffusion and migration becomes significant and dominates the system, overshadowing the charge transfer signal and leading to implausible inferences.

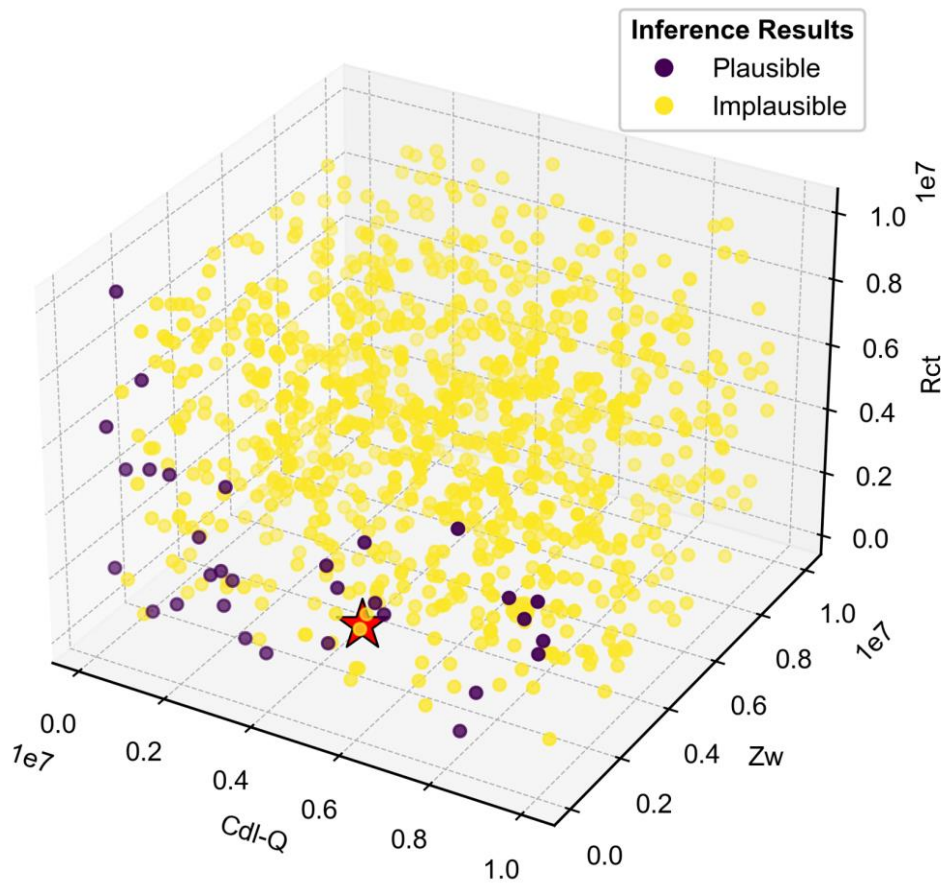

Figure 4. Distribution of statistical plausibility across EIS data with varying reaction parameters for the passivating system. Purple dots: exhibiting well-distributed posterior distributions for all circuit elements; Yellow dots: containing poorly distributed posterior distribution for certain elements. $C_{dl-Q}$: magnitude of double-layer CPE, $Z_w$: Warburg impedance, $R_{ct}$: charge-transfer resistor. Red mark: the selected EIS data used in the subsequent frequency reduction analysis.

Similar to the above, we built a decision tree model using the component values and

BI results, as visualized in Fig. S3. According to the model, the magnitude of the Warburg element ($Z_w$) was identified as the most influential factor, weighing over 85% of the feature importance. For values of $Z_w$ lower than $1.03 * 10^6$ $\Omega \cdot s^{-\alpha}$, over 3.9% of EIS data exhibited plausible inference, while all EIS data encountered failed inferences with a $Z_w$ larger than this value. The second influential parameter is the magnitude of the double-layer CPE ($C_{dl-Q}$). In regions where $Z_w$ is smaller than $1.03 * 10^6$ $\Omega \cdot s^{-\alpha}$ and $C_{dl-Q}$ is larger than $2.89 * 10^6$ $\Omega \cdot s^{-\alpha}$, the BI failed for all EIS data.

Fig 5 illustrates the effect of the frequency reduction on the inference of $Z_w$ using the selected EIS data. The results show that the inferences on $Z_w$ remained both accurate and confident within a 4% relative error compared to the ground truth down to 0.01 Hz. The inference on other ECM components also remains accurate with less than a 4% relative error, as shown in Fig. S4. Beyond this range, the error exceeds 4%, which may still fall within acceptable limits depending on user judgment. Despite the increased uncertainty, the resulting reduction in measurement time is significant (over 90% of the total measurement duration).

The inference maintains its confidence until 0.1 Hz, after which a large number of divergences occur in conjunction with a substantial increase in the relative error, marking a decreased model reliability. Although the posterior distributions of circuit elements remain well-centered, it is recommended to use the ECM with caution for this range of frequencies. For frequencies above 1.0 Hz, inferring the circuit structure becomes impractical with implausible posterior distributions for circuit components.

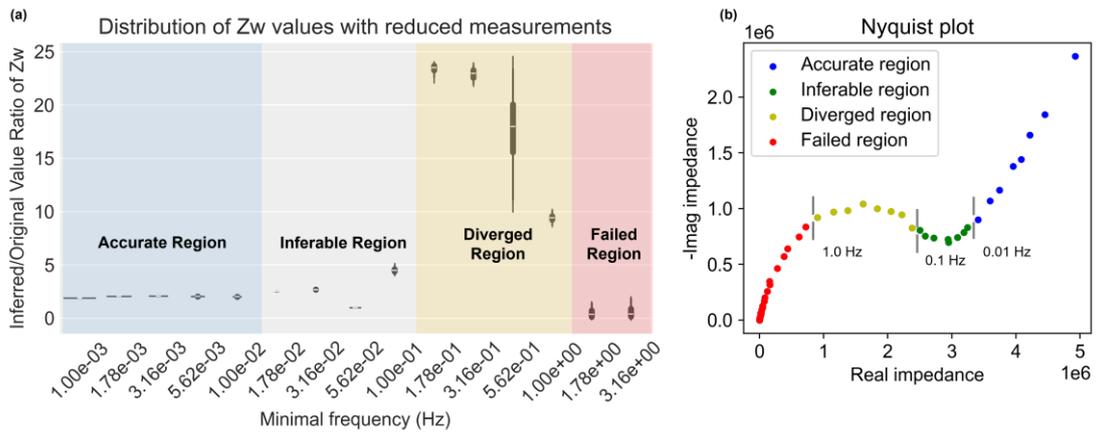

Figure 5. Impact of Reducing Low-Frequency Measurements on the Inference Accuracy for the Selected EIS Data of the Passivation System. (a) Violin plot showing the changes in the selected circuit component value ($Z_w$) as low-frequency measurements are progressively excluded. (b) Nyquist plot illustrating the minimal frequency for regions with different statistical plausibility.

**Coated Metal Corrosion Systems**

The BI results of the coating system are depicted in Figure 6. The data in this figure suggests that a relatively smaller pore resistance ($R_{po}$) is beneficial in obtaining plausible BI results within the specified ECM.

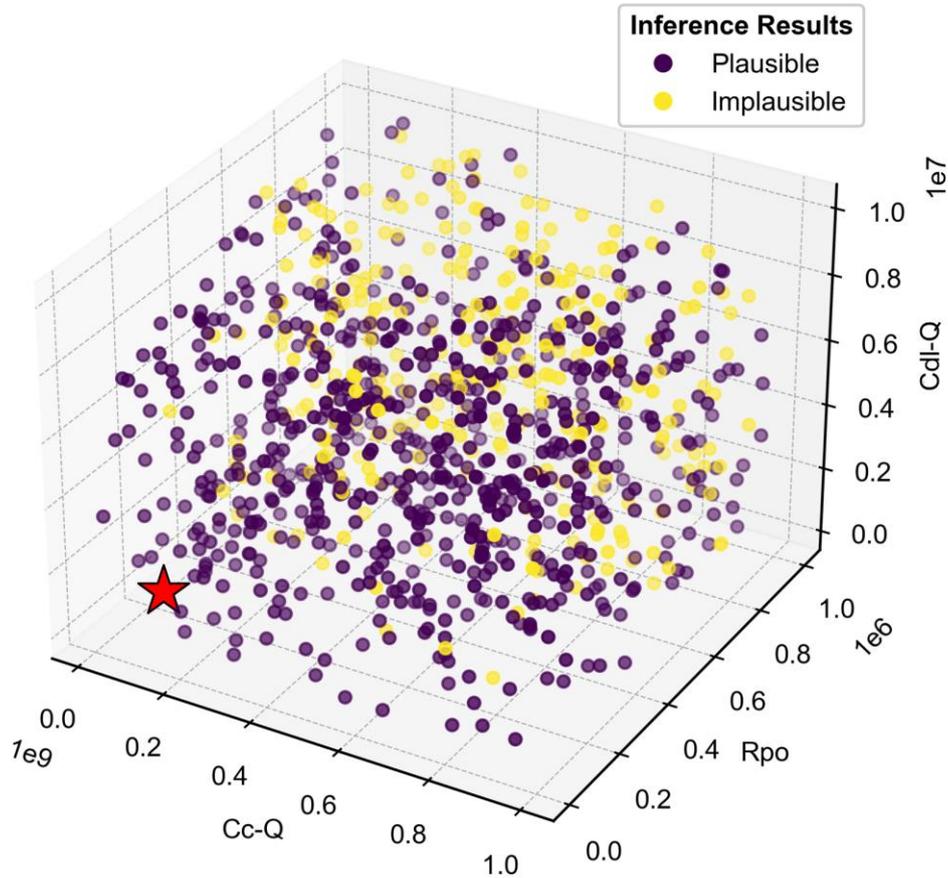

Figure 6. Distribution of statistical plausibility across EIS data with varying reaction parameters for the coating system. Purple dots: exhibiting well-distributed posterior distributions for all circuit elements; Yellow dots: containing poorly distributed posterior distribution for certain elements. $C_{c-Q}$: magnitude of coating CPE, $R_{po}$: pore resistance, $C_{dl-Q}$: magnitude of double-layer CPE. Red mark: the selected EIS data used in the subsequent frequency reduction analysis.

After training a decision tree model using the adjusted parameters and BI results, the pore resistance ($R_{po}$) was identified as the most significant factor for successful inference, accounting for approximately 45% percent of the feature importance. The decision tree results (as shown in Fig. S5) indicate an initial separation when the pore resistance ($R_{po}$) exceeds 4.54 * 10^5 Ω. EIS data with $R_{po}$ below this threshold exhibit plausible posterior distributions in over 89.3% of cases, suggesting sufficient signal to substantiate the given circuit structure.

The influence of frequency reduction on the inference of the magnitude of Warburg element ($Z_w$) is demonstrated in Fig. 7. For the selected EIS which features two semi-circles, omitting low-frequency measurements below 0.018 Hz has a negligible impact on the inference accuracy on $Z_w$ with less than a 4% error but reduces over 94.3% measurement time. Other ECM components also retain this level of accuracy, as demonstrated in Fig. S6. Utilizing BI with impedance data above this threshold, users retain the capacity to confidently infer component parameters while the error exceeds 4%. Furthermore, BI can reliably suggest the ECM structure up to the point where all frequencies below 0.18 Hz are excluded, after which divergence emerges in the inference results and it marks a decreased EIS-ECM pair alignment. After removing measurements below 1.0 Hz, the ECM structure becomes non-inferable.

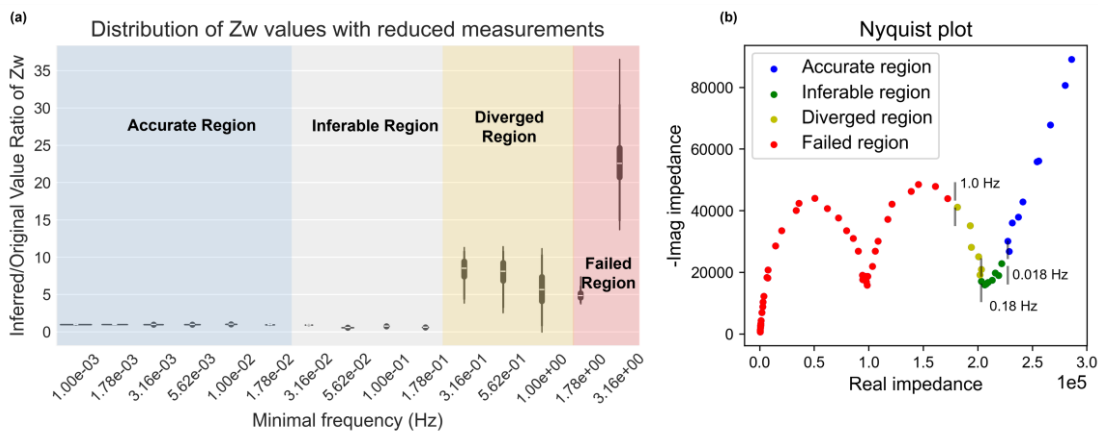

Figure 7. Impact of Reducing Low-Frequency Measurements on the Inference Accuracy for the Selected EIS Data of the Coating System. (a) Violin plot showing the changes in the selected circuit component value ($Z_w$) as low-frequency measurements are progressively excluded. (b) Nyquist plot illustrating the minimal frequency for regions with different statistical plausibility)

**Conclusion:**

Here, we used BI, to demonstrate a framework for evaluating the appropriateness of an ECM based on the information contained in an EIS measurement. Using three common ECMs used to describe corrosion, we identify regions where, despite the fits showing low mean square errors, the credibility of these models is undermined. From this, we propose statistical heuristics for confidently selecting ECMs, which we believe to be generally valid. Additionally, we show that this framework can be used to intelligently select the minimal frequency required to ensure the integrity of analysis, which can substantially reduce the data acquisition time required for accurate EIS analysis. For example, this opens a pathway for rapid in-situ EIS measurements that do not compromise the reliability of analysis. Through this approach, one need only perform a full spectrum frequency measurement on the system's initial state, and subsequent measurements need only be performed down to the critical frequency without risking loss of information. In summary, our work provides a guide toward more informed ECM selection during EIS analysis and proposes a framework to expedite EIS measurements, advancing the application of EIS in in-situ electrochemical degradation.

## Acknowledgments

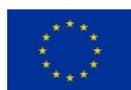This research is part of a project StoRIES that has received funding from the European Union's Horizon 2020 research and innovation programme under grant agreement No 101036910. It was undertaken thanks in part to funding provided to the University of Toronto's Acceleration Consortium from CFREF-2022-00042. The authors gratefully acknowledge the financial support from the National Research Council of Canada under the Canadian Collaboration Center for Green Energy Materials, and the financial support from the Office of Naval Research through the Multidisciplinary University Research Initiative (MURI) program (award #: N00014-20-1-2368) with program manager Dr. D. Shifler.

## Reference

1. Hattrick-Simpers, J. *et al.* Designing durable, sustainable, high-performance materials for clean energy infrastructure. *Cell Rep Phys Sci* **4**, 101200 (2023).

2. Taylor, C. D., Lu, P., Saal, J., Frankel, G. S. & Scully, J. R. Integrated computational materials engineering of corrosion resistant alloys. *npj Materials Degradation 2018 2:1* **2**, 1–10 (2018).

3. Revie, R. W. *Corrosion and Corrosion Control: An Introduction to Corrosion Science and Engineering*. (John Wiley & Sons, 2008).

4. Hou, B. *et al.* The cost of corrosion in China. *npj Materials Degradation 2017 1:1* **1**, 1–10 (2017).

5. Gandoman, F. H. *et al.* Concept of reliability and safety assessment of lithium-ion batteries in electric vehicles: Basics, progress, and challenges. *Appl Energy* **251**, 113343 (2019).

6. Revie, R. W. & Uhlig, H. H. *Corrosion and Corrosion Control: An Introduction to Corrosion Science and Engineering*. *John Wiley & Sons* (John Wiley & Sons, 2008).

7. Li, X. *et al.* Materials science: Share corrosion data. *Nature 2015 527:7579* **527**, 441–442 (2015).

8. Stansbury, E. E. & Buchanan, R. A. *Fundamentals of Electrochemical Corrosion*. (ASM International, 2000).

9. Mansfeld, F. Electrochemical impedance spectroscopy (EIS) as a new tool for investigating methods of corrosion protection. *Electrochim Acta* **35**, 1533–1544 (1990).

10. Wang, S. *et al.* Electrochemical impedance spectroscopy. *Nature Reviews Methods Primers 2021 1:1* **1**, 1–21 (2021).

11. Silverman, D. C. & Carrico, J. E. Electrochemical Impedance Technique — A Practical Tool for Corrosion Prediction. *Corrosion* **44**, 280–287 (1988).

12. Darowicki, K., Krakowiak, S. & Slepski, P. ´. Evaluation of pitting corrosion by means of dynamic electrochemical impedance spectroscopy. *Electrochim Acta* **49**, 2909–2918 (2004).

13. Liu, C., Bi, Q., Leyland, A. & Matthews, A. An electrochemical impedance spectroscopy study of the corrosion behaviour of PVD coated steels in 0.5 N NaCl aqueous solution: Part II.: EIS interpretation of corrosion behaviour. *Corros Sci* **45**, 1257–1273 (2003).

14. Cai, B. *et al.* An experimental study of crevice corrosion behaviour of 316L stainless steel in artificial seawater. *Corros Sci* **52**, 3235–3242 (2010).

15. Macdonald, D. D. Review of mechanistic analysis by electrochemical impedance spectroscopy. *Electrochim Acta* **35**, 1509–1525 (1990).

16. Macdonald, D. D. Why Electrochemical Impedance Spectroscopy is the Ultimate Tool in Mechanistic Analysis. *ECS Trans* **19**, 55–79 (2009).

17. Harrington, D. A. & Van Den Driessche, P. Mechanism and equivalent circuits in electrochemical impedance spectroscopy. *Electrochim Acta* **56**, 8005–8013 (2011).

18. MacDonald, D. D. Reflections on the history of electrochemical impedance spectroscopy. *Electrochim Acta* **51**, 1376–1388 (2006).

19. Ciucci, F. Modeling electrochemical impedance spectroscopy. *Curr Opin Electrochem* **13**, 132–139 (2019).

20. Chang, B.-Y. & Park, S.-M. Electrochemical impedance spectroscopy. *Annual Review of Analytical Chemistry* **3**, 207–229 (2010).

21. Scully, J. R., Silverman, D. C. & Kendig, M. W. *Electrochemical Impedance: Analysis and Interpretation*. (ASTM, 1993).

22. El-Azazy, M., Min, M. & Annus, P. Electrochemical impedance spectroscopy.

23. Taylor, S. R. & Gileadi, E. Physical Interpretation of the Warburg Impedance. *Corrosion* **51**,

(1995).

24. Lai, X., Zheng, Y. & Sun, T. A comparative study of different equivalent circuit models for estimating state-of-charge of lithium-ion batteries. *Electrochim Acta* **259**, 566–577 (2018).

25. Lazanas, A. C. & Prodromidis, M. I. Electrochemical Impedance Spectroscopy─A Tutorial. *ACS Measurement Science Au* **3**, 162–193 (2023).

26. Van Gheem, E. *et al.* Electrochemical impedance spectroscopy in the presence of non-linear distortions and non-stationary behaviour: Part I: theory and validation. *Electrochim Acta* **49**, 4753–4762 (2004).

27. Ciucci, F. & Chen, C. Analysis of Electrochemical Impedance Spectroscopy Data Using the Distribution of Relaxation Times: A Bayesian and Hierarchical Bayesian Approach. *Electrochim Acta* **167**, 439–454 (2015).

28. Barsukov, Y. & Macdonald, J. R. Electrochemical impedance spectroscopy. *Characterization of materials* **2**, 898–913 (2012).

29. Tian, J. *et al.* Simultaneous prediction of impedance spectra and state for lithium-ion batteries from short-term pulses. *Electrochim Acta* **449**, 142218 (2023).

30. Chang, C. *et al.* Fast EIS acquisition method based on SSA-DNN prediction model. *Energy* **288**, 129768 (2024).

31. Li, K., DeCost, B., Choudhary, K., Greenwood, M. & Hattrick-Simpers, J. A critical examination of robustness and generalizability of machine learning prediction of materials properties. *npj Computational Materials 2023 9:1* **9**, 1–9 (2023).

32. Huang, J., Papac, M. & O'Hayre, R. Towards robust autonomous impedance spectroscopy analysis: A calibrated hierarchical Bayesian approach for electrochemical impedance spectroscopy (EIS) inversion. *Electrochim Acta* **367**, 137493 (2021).

33. Zhang, R. *et al.* Editors' Choice—AutoEIS: Automated Bayesian Model Selection and Analysis for Electrochemical Impedance Spectroscopy. *J Electrochem Soc* **170**, 086502 (2023).

34. Kruschke, J. Doing Bayesian data analysis: A tutorial with R, JAGS, and Stan. (2014).

35. Neath, A. A. & Cavanaugh, J. E. The Bayesian information criterion: background, derivation, and applications. *Wiley Interdiscip Rev Comput Stat* **4**, 199–203 (2012).

36. Mansfeld, F. Models for the impedance behavior of protective coatings and cases of localized corrosion. *Electrochim Acta* **38**, 1891–1897 (1993).

37. Mansfeld, F., Shih, H., Greene, H. & Tsai, C. H. Analysis of EIS data for common corrosion processes. *ASTM Special Technical Publication* **1188**, 37 (1993).

38. MacDonald, D. D. Reflections on the history of electrochemical impedance spectroscopy. *Electrochim Acta* **51**, 1376–1388 (2006).

39. Van Haeverbeke, M., Stock, M. & De Baets, B. Practical Equivalent Electrical Circuit Identification for Electrochemical Impedance Spectroscopy Analysis with Gene Expression Programming. *IEEE Trans Instrum Meas* **70**, (2021).

40. Brug, G. J., van den Eeden, A. L. G., Sluyters-Rehbach, M. & Sluyters, J. H. The analysis of electrode impedances complicated by the presence of a constant phase element. *J Electroanal Chem Interfacial Electrochem* **176**, 275–295 (1984).

41. Jewson, J., Smith, J. Q. & Holmes, C. Principles of Bayesian Inference Using General Divergence Criteria. *Entropy 2018, Vol. 20, Page 442* **20**, 442 (2018).

42. Betancourt, M. A conceptual introduction to Hamiltonian Monte Carlo. *arXiv preprint arXiv:1701.02434* (2017).

# An Assessment of Commonly Used Equivalent Circuit Models for Corrosion Analysis: A Bayesian Approach to Electrochemical Impedance Spectroscopy

*Runze Zhang[a], Debashish Sur[b,c], Kangming Li[a], Julia Witt[d], Robert Black[e], Alexander Whittingham[e], John R. Scully[b,c], Jason Hattrick-Simpers[a]\**

- a. Department of Material Science and Engineering, University of Toronto, Toronto, Ontario, Canada
- b. Center for Electrochemical Science and Engineering, University of Virginia, Charlottesville, VA USA 22904
- c. Department of Materials Science and Engineering, University of Virginia, Charlottesville, VA USA 22904
- d. Division of Material and Surface Technologies, Federal Institute of Materials Research and Testing (BAM), 12205 Berlin, Germany
- e. Clean Energy Innovation Research Centre (CEI), National Research Council Canada, Mississauga, Ontario, Canada

Email: jason.hattrick.simpers@utoronto.ca

**Supporting Information:**

1. **Charge-Transfer Controlled Homogenous Surface Corrosion Analysis**

   The decision-making processes within the Decision Tree model trained on the generated homogeneous corrosion EIS dataset are visualized in Fig. S1.

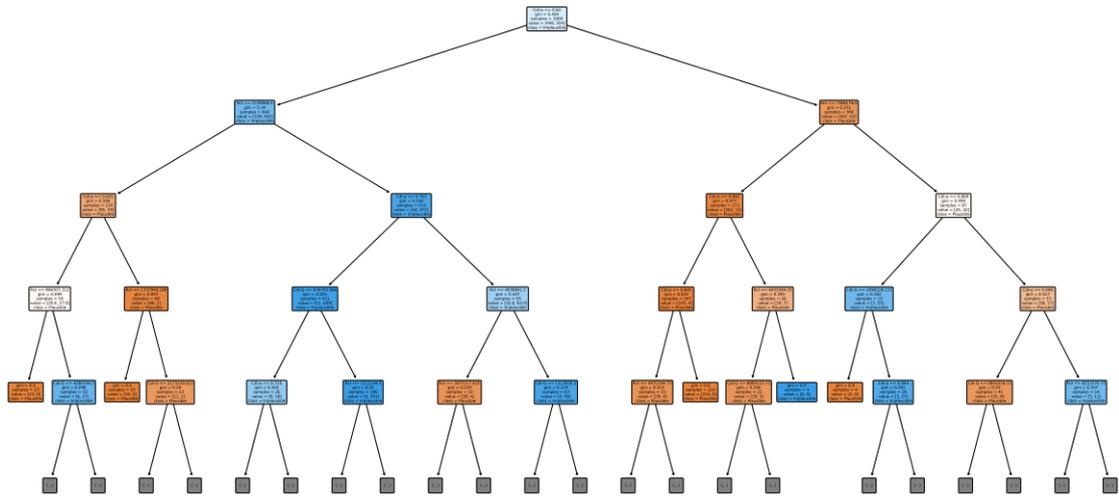

Figure S1. Decision-making processes of the decision tree model trained on the uniform corrosion EIS dataset

Fig. S2 illustrates the impacts of reducing the minimum measured EIS frequency on the inference of all ECM components for the uniform corrosion system. As the frequency reduction proceeds, the posterior distributions of all ECM components broaden, indicating decreased confidence. Additionally, the center of the inferred posterior distributions deviates from the ground truth values used to generate the EIS data.

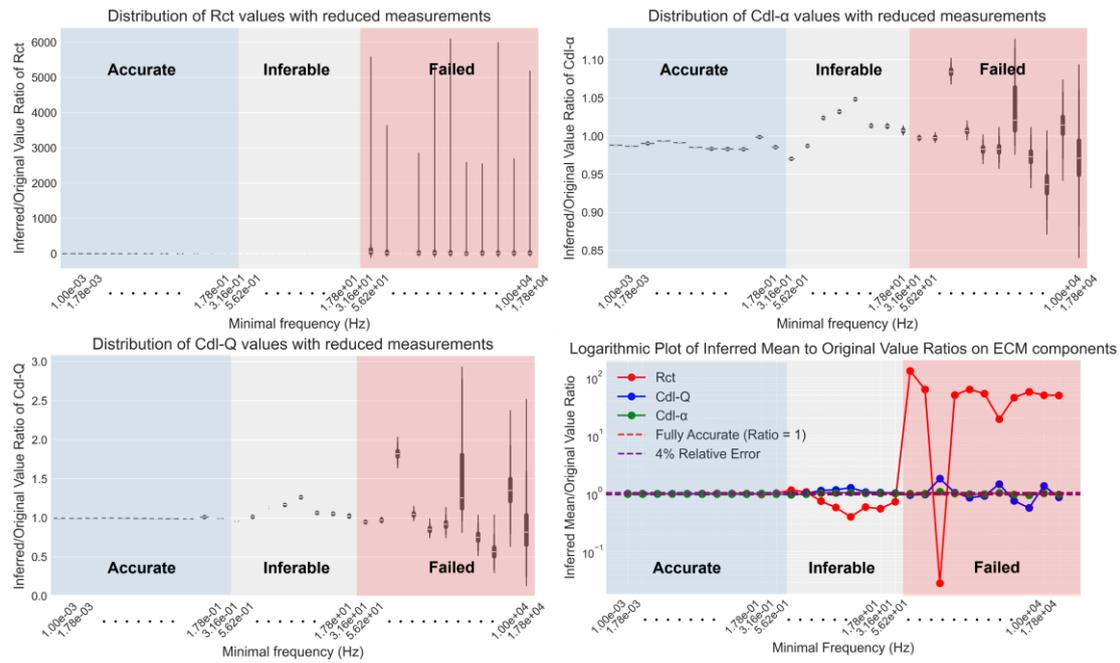

Figure S2. The impact of reducing the minimum measured EIS frequency on ECM components for the uniform corrosion system. (a) $R_{ct}$ (b) $C_{dl-Q}$ (c) $C_{dl-\alpha}$ (d) The ratios of the mean of the inferred posterior distributions to the ground truth component values on ECM components.

## 2. Passivation of Alloys Under Aqueous Condition

The Decision Tree model trained on the passivating system EIS dataset is shown in Fig. S3.

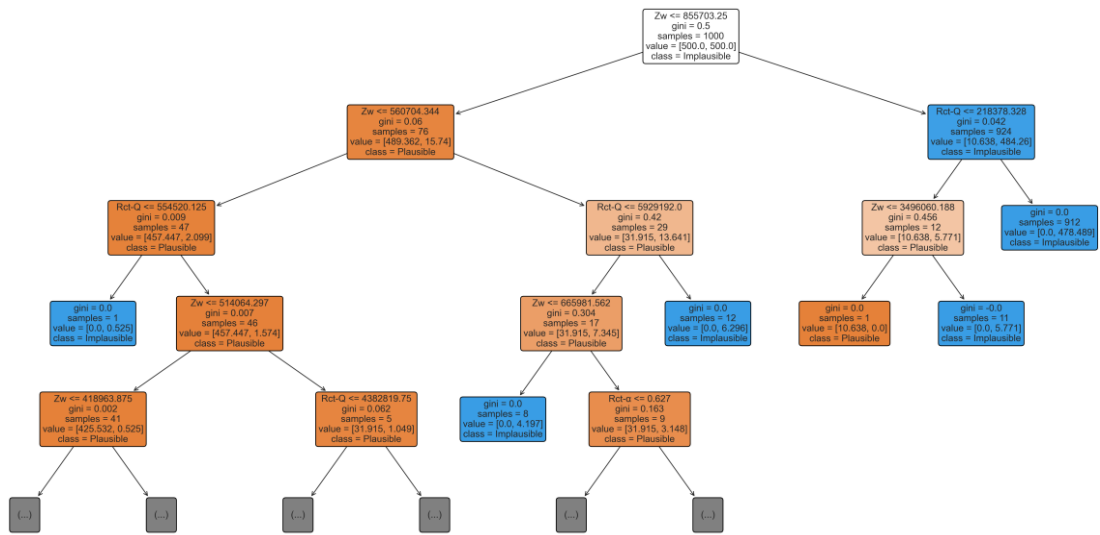

Figure S3. Decision-making processes of the decision tree model trained on the passivating EIS dataset

Fig. S4 demonstrates the variations on the inference of ECM components as the frequency measurement reduction continues.

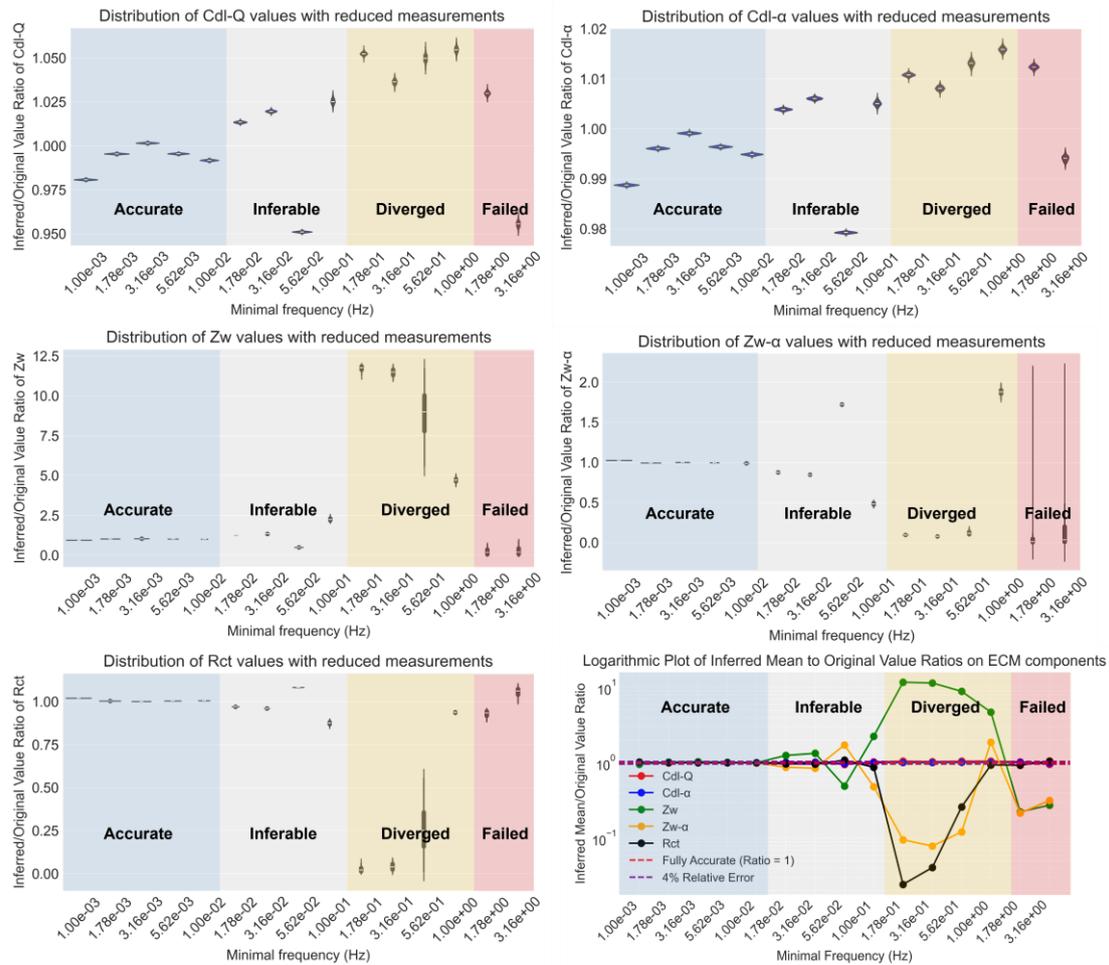

Figure S4. The impact of reducing the minimum measured EIS frequency on ECM components for the passivating system. (a) $C_{dl-Q}$ (b) $C_{dl-\alpha}$ (c) $Z_w$ (d) $Z_{w-\alpha}$ (e) $R_{ct}$ (f) The ratios of the mean of the inferred posterior distributions to the ground truth component values on ECM components.

## 3. Coated Metal corrosion systems

The Decision Tree model trained on the coating system is shown in Fig. S5.

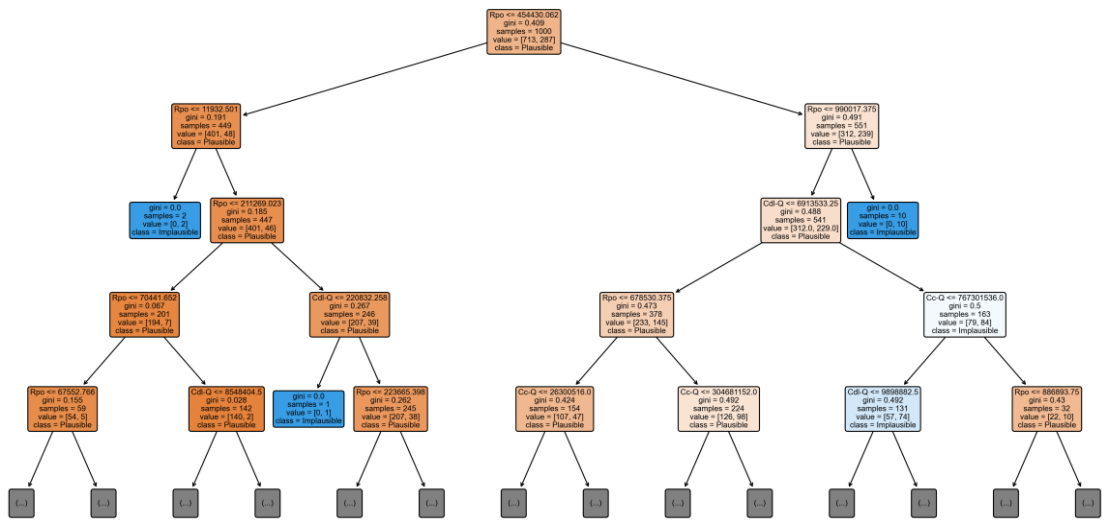

Figure S5. Decision-making processes of the decision tree model trained on the coating EIS dataset

Fig. S6 demonstrates the influence of reducing the minimum measured EIS frequency on the inference of all ECM components for the coating system.

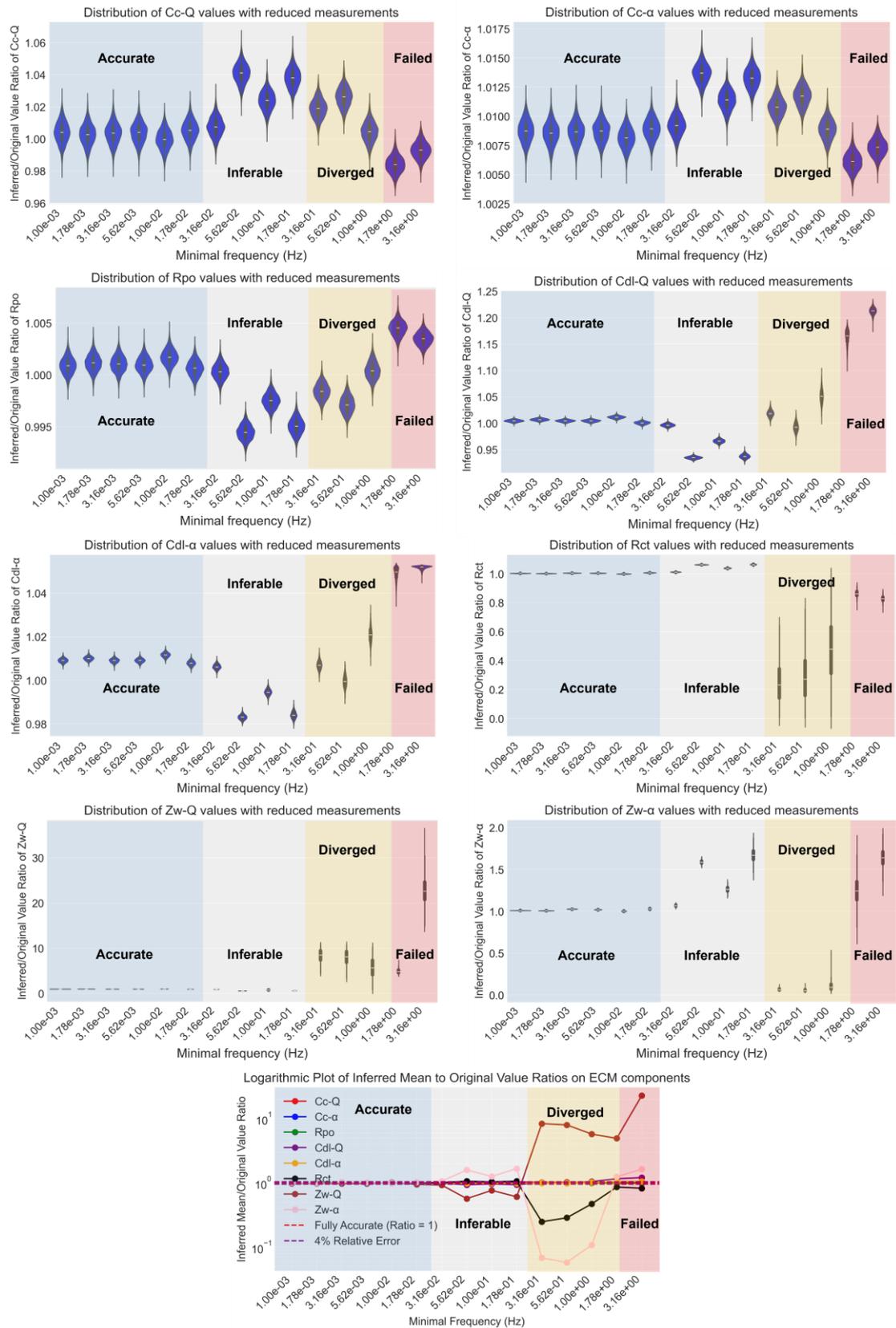

Figure S6. The impact of reducing the minimum measured EIS frequency on ECM components for the coating system. (a) $C_{c\text{-}Q}$ (b) $C_{c\text{-}\alpha}$ (c) $R_{po}$ (d) $C_{dl\text{-}Q}$ (e) $C_{dl\text{-}\alpha}$ (f) $R_{ct}$ (g)

$Z_{w\text{-}Q}$ (f) $Z_{w\text{-}\alpha}$ (g) The ratios of the mean of the inferred posterior distributions to the ground truth component values on ECM components.

## 4．Influences of noise on the accuracy and confidence thresholds

We demonstrate how different levels of noise influence the ECM accuracy and confidence thresholds by showing the inference results with two selected levels of noise, 2% and 0.1%, as shown in Fig. S7. With a smaller level of noise, both the accuracy and confidence threshold got postponed, indicating the possibility of omitting more low-frequency measurements without detriment of the integrity of EIS analysis. The accuracy threshold increases from 0.01 Hz to 0.056 Hz, and the confidence threshold increases from 0.1 Hz to 1.0 Hz. Moreover, the diverged region becomes narrower, marking increased confidence in the given ECM.

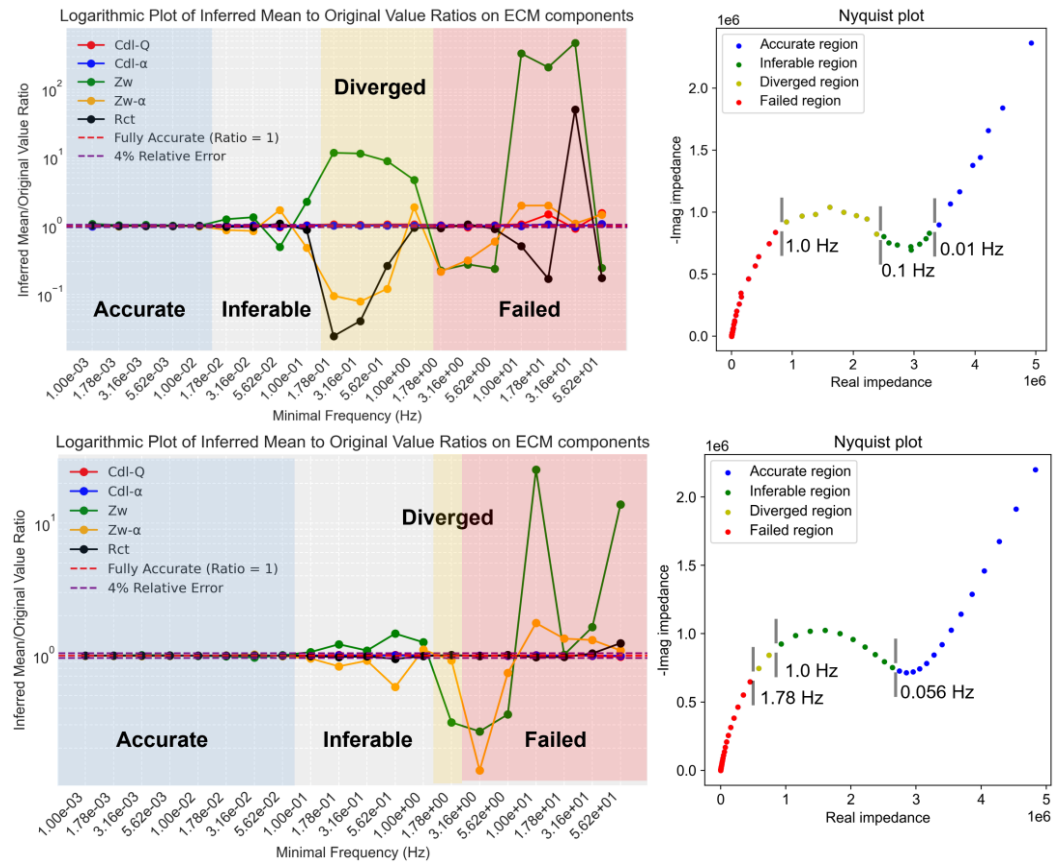

Figure S7. Impacts of reducing low-frequency measurements on the inference accuracy and confidence at different noise levels (a) 2% and (b) 0.1%